# Visualizations in Exploratory Search – A User Study with Stock Market Information

Daniel Hienert and Philipp Mayr
GESIS – Leibniz Institute for the Social Sciences
Unter Sachsenhausen 6-8, 50667 Cologne, Germany
+49 - (0)221 – 47694 - 0

{daniel.hienert, philipp.mayr}@gesis.org

## ABSTRACT
In this paper we present an approach that integrates interactive visualizations in the exploratory search process. In this model visualizations can act as hubs where large amounts of information are made accessible in easy user interfaces. Through interaction techniques this information can be combined with related information on the World Wide Web. We applied the new search concept to the domain of stock market information and conducted a user study. Participants could use this interface without instructions, could complete complex tasks like identifying related information items, link heterogeneous information types and use different interaction techniques to access related information more easily. In this way, users could quickly acquire knowledge in an unfamiliar domain.

## Categories and Subject Descriptors
H.3.3 [**Information Storage and Retrieval**]: Information Search and Retrieval – Search Process

## 1. INTRODUCTION
The explorative search model is a good abstraction for the information seeking process in the World Wide Web. Users browse between different media types like web pages, images, photos, videos and visualizations to solve an information need while using different search techniques. In the search process users learn from the processed information and acquire knowledge for the following search step. Visualizations from simple info graphics to complex interactive visualizations can be one step in the search process, whereby the visualization of information can have many advantages. Large amounts of data and information can be presented in graphics and can be accessed and explored with interaction techniques. Based on the visualization type different facets of information can be explored more easily and users can discover patterns and trends.

However, so far, interactive visualizations on the web are barely used in an overall explorative information seeking process. Either, they are representing information from one or more sources and interaction techniques can be used to explore this information inside the (coordinated) visualization(s), or they are used as widgets to filter the information items they are representing. But, underlying information items often have strong or weak connections to other related information items in the web and these relations can be used in an overall search process for learning and knowledge acquisition.

Based on this approach we have designed a system that shows stock data and financial news in coordinated visualizations with links to online articles. Users can explore data and find relations between information items to link them and use interaction techniques. We have conducted a user study to test our approach and found that users can acquire knowledge in an unfamiliar domain.

The paper is organized as follows: Section 2 presents related work in the fields of information seeking and information visualization, Section 3 introduces a search concept as a combination of interactive visualizations and explorative search. Section 4, 5, 6 and 7 presents a user study where the model is applied to the domain of financial information. We will conclude in Section 8.

## 2. RELATED WORK
Related work as a basis for our work can be found in research fields like information seeking and information visualization.

### 2.1 Searching and Exploring Information
In the *classical IR model* [4] the information need of a user is represented in a query and compared with the surrogates of documents from a database. A broader view of the search process is offered by the *berrypicking* model from Bates [3]. In this concept information seeking is a dynamic process in which the user applies different search techniques to browse between documents and information. Based on information bits a learning process lets constantly evolve the information need and the search query is adapted. *Exploratory search* [14, 22] extends the berrypicking approach and applies the search process as overlapping activities of lookup, learn and investigate. It is focused on a learning process that includes multiple iterations to cognitively process different information and to create new knowledge. The concept of *information scent* [7, 23] tries to predict the navigation behavior of a user while browsing the web on the basis of proximal cues, such as web links, icons and surrounding text. At each navigation step, the user weights value and costs and decides whether to follow this path or not. *Facetted Navigation* [9] is a popular concept that makes information spaces inferable using orthogonal sets of metadata. The user don't need to enter any queries, but can browse by automatically created categories or can combine characteristics to filter search results.

## 2.2 Interactive Information Visualization

A lot of systems allow the publication of information with *visualizations on the web*. Applications like IBM Many Eyes [20] allow the user to upload their own data and share it with the community. Other toolkits are used to create dashboards with visualizations. Tableau [13] or Spotfire [1] are professional solutions which allow the creation of dashboards with different visualizations. Dashiki [15] is a wiki-based collaborative platform for creating visualization dashboards. Users can integrate visualizations that contain live connections to data sources. Exhibit [12] is a lightweight framework for easy publishing of structured data on the web. Users can import data and present it in different views like maps, tables, thumbnails and timelines. Different *interaction techniques* are used to filter data or to find relations among data in visualizations. *Dynamic queries* [2] allow the filtering of data with easy-to-use sliders. Following the design mantra of Shneiderman, data is first shown in an overview and the user can interactively zoom and filter to details of interest. *Coordinated Multiple Views* [17, 21] shows the same or related data sets in different views on one page. Each view uses different visualization types to visualize different characteristics of the data. Additionally, different interaction techniques can be applied. For example, *brushing-and-linking* highlights the same data in all views, if it is selected in one view. The Vizgr [11] toolkit introduces the approach of *linking data* directly in visualizations and using *interaction icons* in visualizations for browsing, querying and filtering from visual elements. *Searching and filtering of web information* is an important new role of interactive visualizations. For example, the system VisGets [8] uses different visualizations to show and filter retrieved web resources in several dimensions like time, location and topic. Based on the concept of dynamic queries, results can interactively be filtered by manipulating the visualizations. Other prominent visualization types in the web like tag clouds and maps have been used for several information seeking tasks [19, 24].

In the following section we show the benefits of the integration of interactive visualizations in the explorative search process.

## 3. VISUALIZATIONS IN EXPLORATORY SEARCH

Bates [3] argues for a search model that is closer to the real search behavior of information seekers. Users browse between different information and documents using different search techniques to learn from them and adapt the information need until the satisfactory completion of the search task. Marchionini refers more to the learn and investigate step. Multiple iterations and cognitive processing and interpretation of objects in various media like graphs, maps, texts and videos is needed. The user has to spend time "scanning/viewing, comparing and making qualitative judgments" for the result of "knowledge acquisition, comprehension of concepts or skills, interpretation of ideas, and comparisons or aggregations of data and concepts" [14]. Very similar, Information Visualization wants to support large quantities of complex information and to achieve insight by discovering patterns, relationships etc. in interactive visualizations [16]. The knowledge crystallization concept [6] is a loop of information foraging, schema creation, problem solving and learning/acting, whereby in each step visualizations can be used.

Interaction techniques allow the integration of visualizations in exploratory search, not only as separate closed systems, but with linkages between information items. The connection allows *browsing* links between information on the user interface. Figure 1 shows the idea of interactive visualizations in the exploratory search process: (1) Visualizations can be used as a gateway to different information types like i.e. stock data or financial news. Information is shown in an overview and can easily be explored. (2) Linkings on the data level and interaction techniques on the user interface allow the user to browse between these heterogeneous information types. (3) Links between information types in visualizations and in the web can be used in an exploratory search process. Based on an information need users can make several iterations over related information using interaction techniques, can learn from processed documents, patterns and relationships and can acquire knowledge.

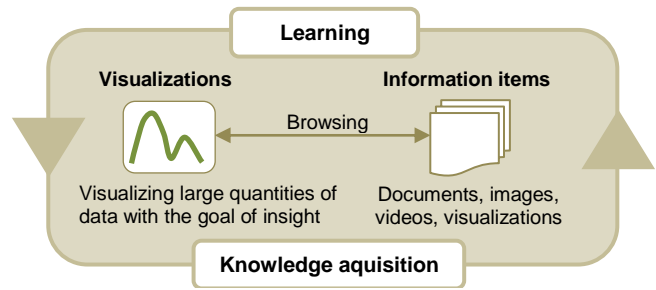

**Figure 1. Integrating interactive visualizations in the exploratory search process by linking related information items.**

In the next section, we will describe the evaluation environment which has been created to apply and evaluate this concept in the domain of financial information.

## 4. EVALUATION ENVIRONMENT

Based on the Vizgr Toolkit [11] an evaluation environment has been created that consists of the following components: (1) a FTSE 100 index chart, (2) a timeline with financial news and (3) a mapping form. Figure 2 shows the evaluation environment.

### 4.1 FTSE 100 Index Chart

The FTSE 100 index is the most important British share index including the one hundred biggest and highest capitalized companies listed on the London Stock Exchange. Data for the index chart has been imported from Yahoo Finance[1], which provides historical prices for many indexes and stocks. The original FTSE 100 data set contains over 7,000 daily prices from 1984 to 2012 with different attributes like *date, open*, *high*, *low*, *close*, *volume* and *adjusted close*. The last field is adjusted for dividends and splits. For the index chart we use only the attribute *adjusted close* and data from 2005 to 2012 with 1,770 values to keep the visualization as simple as possible. The line chart shows daily closing prices in British Pounds from 2005 to present. The diagram is interactive: (1) the user can move the mouse over data points and the mouse-over shows data and the actual price, (2) the graph can be filtered using a dynamic slider to a specific period of time, and (3) a click on a data point loads financial news for that day in the timeline.

### 4.2 Financial News in the Timeline

The timeline shows financial news from the online edition of the daily newspaper *The Guardian*. Data is provided from the

---

[1] http://finance.yahoo.com/q?s=^FTSE

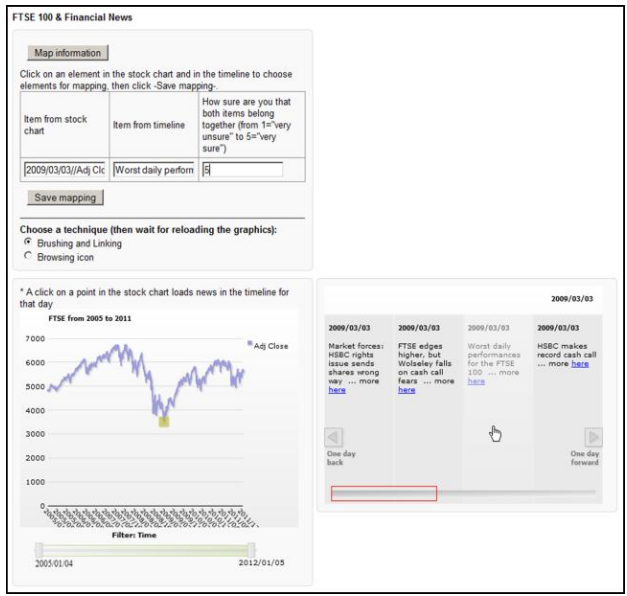

**Figure 2. Evaluation environment with index chart, timeline with financial news and forms for the mapping of information and the selection of interaction techniques.**

Guardian Open Platform[2] that allows querying the article database with keywords. To use the API with the Vizgr toolkit a wrapper was created that is requesting articles on a particular date, limited to the category "economy" and filtered to the keyword "ftse". The wrapper transforms API results into the appropriate XML schema. The API returns metadata for matching articles in JSON format that match these conditions. Each entry consists of a title, publication date, URL, summary and thumbnail. From this data a timeline is created that shows financial news for a day with title, a link to the online article and a thumbnail. Users can move the timeline to browse in financial news for one day or can use buttons on the left and right to move a day backward or forward.

### 4.3 Mapping Form
Participants can combine information items from the index chart and the timeline in the mapping form. This is the prerequisite for the two interaction techniques *brushing-and-linking* and *interaction-icon*. By clicking on the button "*Map information*", the mapping form can be extended (cf. Figure 2). The form has three fields for (1) the identifier of a data point in the index chart, (2) for the identifier of a financial news in the timeline, and (3) for a confidence value. Users can fill out the fields (1) and (2) with a mouse click in the visualizations. Clicking on a data point in the index chart enters the identifier in the left field, clicking on a financial news enters the identifier in the middle field. The user can also enter a confidence score in the third field with the following label: How sure are you that both elements belong together (from 1 = "very unsure" to 5 = "very sure"). By clicking on the button "*Save mapping*" the mapping is applied and the graphics are reloaded. The users can then select between the two interaction techniques *brushing-and-linking* or *interaction-icon* (see Figure 3).

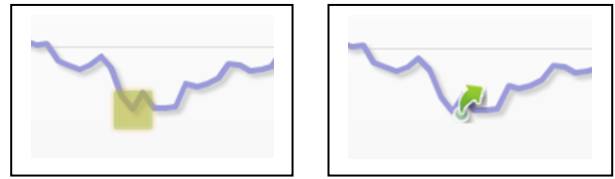

**Figure 3. Highlighting the data point in the graph with brushing-and-linking or with the interaction-icon**

### 4.4 User Interaction Process
The chart shows closing prices of the FTSE 100 index from 2005 to 2012. With the help of the chart, the user can identify index patterns, trends, maxima and minima, can filter down to a certain time period and can analyze data on the details pane. Hovering over a data point shows the date and price for that particular day. By clicking on a data point financial news are loaded in the timeline for that day. The user can scan the titles to explore if there are financial news with associations to the price development. With a mouse click on buttons in the timeline the user can also examine news from previous or following days. If the user has discovered a suitable message corresponding to his information need, he can click on the link to browse to the online article and read details there.

If price development and financial news have a significant correlation for the user, the information items can be connected. The mapping is the basis for the interaction techniques brushing-and-linking and interaction-icon that can facilitate the subsequent interaction in the visualizations. The mapping is defined by clicking on the information elements and described with a confidence value.

In the technique brushing-and-linking associated information elements are highlighted in both views. Hovering with the mouse over the data point in the chart highlights the news and vice versa, hovering over the financial news highlights the data point. This way, it can be recognized intuitively that there is a connection between the elements. This would help end users of a system to see directly a relation between financial news and a price development and one could browse from the timeline to the online article. In a final system with many links, additionally the trust value can be visualized by different shades of color.

In the technique interaction-icon an arrow icon will be displayed directly in the index chart at the data point. By clicking on the icon the user can directly browse to the online article. This way, the process step via the timeline is avoided and thus a cognitive step is saved. However, context information is missing that is given by financial news on the same, previous or next days. Figure 4 shows the possible transitions between index chart, timeline and online articles.

### 5. USER STUDY
We have conducted a user study to verify the following research questions:

(1) Can users work with heterogeneous information types in visualizations and related information on the web in an exploratory search process? Specifically: can relations between price trends and financial news be found intuitively?

---

[2] http://www.guardian.co.uk/open-platform

(2) Does the transition between different types of information and presentation of information such as index chart, timeline and articles on web pages work smoothly?
(3) Can users connect various types of heterogeneous information and based on this mapping can they intuitively use interaction techniques?
(4) Do users perceive interaction techniques as an added-value?
(5) Which technique would users prefer in this application context?

## 5.1 Method
Participants were asked to carry out a variety of tasks in the evaluation environment and to complete a questionnaire after each task with the found response, time required, difficulty level and comments. In a first step, the participants could familiarize themselves with the evaluation environment and the interaction elements in the visualizations for two minutes. In the following step, they had to deal with four tasks in the evaluation environment and answer questions in the questionnaire. After performing the tasks users could evaluate the pros and cons of interaction techniques and assess the overall scenario.

## 5.2 Participants
In Group 1 eleven male researchers aged 27 to 50 years (mean: 35 years) have participated. Of these, nine participants had a graduate degree in computer science, one a master degree in computer science and one a Master of Arts in history. Participants were asked to rate (from 2 = "very good" to -2 = "very bad") their experience in dealing with information portals and information graphics on the internet on a five-point-scale. The average rating was good (1.00) with a standard deviation of 0.82.

Participants in Group 2 were eleven students of business information systems at age 22 to 32 years (mean: 26 years). We paid each student 15€ to proceed tasks online in the test environment and to answer questions in the online survey tool. Three female and eight male students from fourth to tenth semesters (second to fifth year) took part. The self-assessment on the experience with information portals and information graphics on the Internet was normal (0.45) with a standard deviation of 0.69.

## 5.3 Tasks and Questions
Participants had to handle the following tasks and answer questions within the evaluation environment:

*1. Load financial news for index prices*

Which financial news were published on the FTSE 100 index minimum on March 3rd, 2009? Which are published the next day? Note the first three words of the first news of each day.

*2. Find reasons for the rise in the market*
- Find two financial news items in the timeline that explain the rise from the index minimum on 03/03/2009. Therefore, browse in the timeline also until the 10/03/2009 and scan the short news.
- Have you found two short news items, then find additional details in the online article of *The Guardian*.
- How confident are you that the price rise and the financial news are related? Note the confidence value in the table.

*3. Link information*

You can link the minimum rate on 03/03/2009 with a financial news item. Click on the button "Map information". Click on the data point on 03/03/2009 in the index chart and a suitable news item in the timeline. Then click "Save mapping".

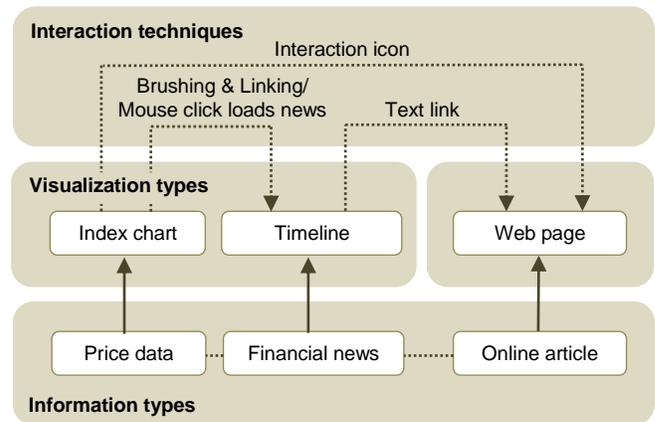

**Figure 4. User interaction process: possible transitions between information types and their visualizations in the index chart, timeline or web page.**

*4. Use interaction techniques on the basis of the mapping*

Based on the mapping, you can now use various techniques in the graphics and switch between them. Try the different techniques.

In task (1) financial news are loaded interactively to prices and the connection between the graphics and information is shown to the user. Task (2) extends the transition to online articles on the web and lets the user search for intellectual connections between the types of information. Task (3) is used to link different types of information interactively and task (4) shall apply interaction techniques based on the mapping. Subsequently, the overall scenario will be evaluated.

For each task from (1) to (4) users had to write down the response in a text field, answer how long the process took, assess as how difficult the task was perceived on a five-point-scale (2=very easy, 1=easy, 0=neutral, -1=difficult, -2=very difficult) and could give comments and suggestions. In task (2) and (3) a confidence score on a five-point-scale (1=very unsure, 2=unsure, 3=normal, 4=sure, 5=very sure) had to be entered. Task (4) contains five-point-scales for the assessment of the techniques brushing-and-linking and interaction-icon. The overall scenario could be rated with free text and a five-point-scale (2=very helpful, 1=helpful, 0=normal, -1=not helpful, -2=not helpful at all) and left room for general comments and suggestions.

## 5.4 Additional Challenges in the Study
There have been some additional challenges for participants in the user study like:

- User interface, financial news and online articles were in English language, specifically in financial English, while all participants of the study were native German speakers.
- Participants from Group 1 were unfamiliar to the domain of finance information and have no professional domain knowledge; participants from Group 2 had at least a slight relation due to their studies.
- For all participants the overall interaction process was fairly new (like linking information in graphics, use interaction techniques, temporal filtering).

# 6. RESULTS

In this section we present the results of the user study organized by tasks:

1. *Load financial news for index prices*

Index prices are displayed in the stock chart, financial news in the timeline. The legend of the index chart contains the information that clicking on a data point in the chart loads financial news for that day in the timeline. The date 03/03/2009 in the chart is initially not accessible with a mouse-over. Following the Mantra of Shneiderman "overview first, zoom and filter, then details on demand" [18], all data is displayed from 2005 to 2012. The user now has to filter the time period with the dynamic slider so that the mouse-over appears for 03/03/2009 and is reachable in a comfortable zoom level. By clicking on the data point appropriate financial news are loaded into the timeline. By clicking on the button "one day forward" in the timeline news for the next day are displayed.

Group 1: ten out of eleven participants were able to answer the question and indicated the correct titles of the financial news. One participant had overlooked the legend, which pointed out that a mouse click on the graph loads financial news for that day and therefore could not answer tasks 1 to 3. The average time for solving the task was 64 seconds with a standard deviation of 51 seconds. The difficulty of the task was classified as normal (0.27). In the comments, it was noted that (1) it was not advised that one can click on the chart to load financial news, (2) the filtering of the slider is inaccurate, and (3) that the data of both graphics are not filtered simultaneously. For Group 2, the temporal slider has been positioned below the stock chart.

Group 2: seven out of eleven participants gave the correct answer, four participants reported news of other days. The average time for achieving the task was 134 seconds with a standard deviation of 106 seconds. The difficulty of the task was classified as normal (0.1). Also in this comments section of this group the inaccuracy of the time slider was noted, and additionally the possibility to offer a calendar or a search box to find news items in the timeline was suggested.

2. *Find reasons for the rise in the market*

After matching financial news have been loaded into the timeline for the 03/03/2009, the user should scroll through the timeline and find appropriate messages that can justify the price increase. Information basis on this level are only the headlines that summarize the article. In the first step users should find two potential news and in a second step justify their choice further by checking details in the online article. The confidence value indicates the confidence of the user in the financial news as a reason for the price rise.

*Group 1:*

Part a (1st financial news): ten out of eleven participants have selected a news item to justify the price increase. Selected messages are shown in Table 1. Nine out of eleven participants were able to cite details from the online articles, so they have selected the link in the timeline and have switched to the online article. The average confidence value for all financial news was *sure* (4.0) with a standard deviation of 0.82.

Part b (2nd financial news): eight out of eleven participants also indicated a second message for justifying the price increase. Seven out of eleven participants also gave details from the online articles. The average confidence value for all financial news was *sure* (3.88) with a standard deviation of 0.99.

**Table 1. Financial news that has been assigned to the price increase (ØC = Average confidence value, C=Confidence value)**

|  | Group 1 | Group 2 |
|---|---|---|
| "Oil stocks help FTSE 100 regain early losses" (09/03/2009) | 7 times ØC: sure (3.71) | 5 times ØC: sure (3.8) |
| "Bank of England ready to pump money into UK economy" (05/03/2009) | 4 times ØC: sure (4) | 2 times ØC: normal (3) |
| "Nick Fletcher: Ray of light from China shines on FTSE" (05/03/2009) | 3 times ØC: very sure (5) | 3 times ØC: sure (3.67) |
| "Market Forces: FTSE rises on hopes of Opec production cuts" (10/03/2009) | 2 times ØC: normal (3) | 3 times ØC: normal (2.67) |
| "Shares bounce back in Asia and London" (04/03/2009) | 1 time C: very sure (5) |  |
| "Citigroup chief and Bernanke lift world markets" (10/3/2009) | 1 time C: normal (3) |  |
| "Financials give a fillip to the FTSE 100" (10/03/2009) |  | 2 times ØC: sure (3.5) |
| "Miners help FTSE 100 to an early rebound" (04/03/2009) |  | 1 time C: unsure (2) |
| "FTSE edges higher, but Wolseley falls on cash call fears" (03/03/2009) |  | 1 time C: unsure (2) |
| "FTSE 100 jumps nearly 5% as financial shares rebound" (10/03/2009) |  | 1 time C: very sure (5) |
| "China stimulus hopes push up world stockmarkets" (13/03/2009) |  | 1 time C: sure (4) |
| „Market forces: Insurers push FTSE 100 back down hill" (06/03/2009) |  | 1 time C: normal (3) |
| „US bank bailout plan sends FTSE 100 soaring" (23/03/2009) |  | 1 time C: normal (3) |

The average time to solve the complete task was 227 seconds with a standard deviation of 186 seconds. The perceived difficulty of the task was classified as difficult (-0.55). In the comments, it was noted that the financial English was hard to understand.

*Group 2:*

Part a (1st financial news): all participants have selected financial news to reason the price increase. All participants were able to cite details of the online articles. The average confidence value of all financial news was sure (3.56) with a standard deviation of 0.98.

Part b (2nd financial news): also here all participants gave a second message to justify the price increase. All participants also gave details from the online articles. The average confidence value of all financial news was normal (3.04) with a standard deviation of 0.51. The average time to solve the complete task was 360 seconds with a standard deviation of 172 seconds. The difficulty of the task was classified as difficult (-0.82).

*3. Link information*

Now, users could link the index minimum with an appropriate financial news item, assign a confidence value and save the mapping.

*Group 1:* all participants were able to solve the task. The participants chose different news to link to the index minimum as shown in Table 2. The average time to solve the complete task was 42 seconds with a standard deviation of 15 seconds. The difficulty of the task was classified as normal (0.64).

*Group 2:* again here all participants were able to solve the task. One user has not entered correctly the financial news into the questionnaire, the other participants chose various news of the 03/03/2009. The average time to solve the complete task was 139 seconds with a standard deviation of 93 seconds. The difficulty of the task was classified as normal (0.37).

*4. Use interaction techniques on the basis of the mapping*

*Group 1:* only six out of eleven participants were able to explain the effects of the technique brushing-and-linking. The comments suggest that (a) the mouse-over area on the diagram (only one data point) is too small and cannot be found directly and (b) the visual highlighting on the chart was too small and overlooked. Ten out of eleven participants were able to explain the technique interaction-icon, identified the icon in the chart and could explain the functionality. As the benefits of brushing-and-linking were noted, that it is well suited for many links, but for only a few links it is difficult to recover the data points in the chart. As the advantages of the technique interaction-icon was noted, that there is a direct link to the article, important points in the chart are always visible and you can use it as a personal marker or bookmark. As a disadvantage it was noted that for many links the use of the icon would be too confusing, because there are too many points in the chart that are marked with the icon. The interaction technique brushing-and-linking was evaluated with normal (-0.33, standard deviation: 1.58), the technique interaction-icon with useful (0.56, standard deviation: 1.13).

*Group 2:* six out of eleven people were able to explain the effect of the brushing-and-linking technique. Problems from Group 1 (visual highlighting to small) have been avoided. The comments stated that no difference could be noticed after the selection of the technique. Most of the participants have probably not studied further the graphics, and without interaction with the mouse no optical effect could be seen. Nine out of eleven participants explained the effect of the interaction-icon, identified the icon in the chart and could explain the functionality. As advantages for brushing-and-linking was noted that the correlation or relationship between data is made clear. As advantages of the interaction-icon was noted that important items are accessible by a direct link, and having a bookmark effect, since data in the chart can be found quickly. A disadvantage of the interaction-icon is that the correlation between data is not visible. The brushing-and-linking technique was classified normal (-0.09, standard deviation: 0.94), the technique interaction-icon with useful (0.73, standard deviation: 0.65). Three participants chose brushing-and-linking as the preferred technology for this scenario, six chose interaction-icon.

*Final assessment*

*Group 1:* overall, users found the connection between index prices and financial news helpful (0.57, standard deviation: 0.98) for the analysis of index data. In general comments was noted that knowledge in economic and financial English is helpful for the analysis and that further explanations for the interaction techniques would be helpful and the connections between the graphics would have to be made more explicit.

*Group 2:* also here users found the connection between index prices and financial news helpful (0.91, standard deviation: 0.83). The user stated in the comments that it was certainly helpful to have multiple data sources for the assessment of prices at a glance, to see the context of the news and get possible links highlighted. The users also noted that other factors certainly play a role for the price development. In the general comments were stated, that considerably more information should be displayed. Table 3 presents the summarized results.

**Table 2. Financial news that has been assigned to the FTSE index minimum (ØC = Average confidence value, C=Confidence value)**

|  | Group 1 | Group 2 |
|---|---|---|
| "Worst daily performances for the FTSE 100" (03/03/2009) | 6 times ØC: sure (3.66) | 4 times ØC: sure (4) |
| "Market forces: HSBC rights issue sends shares wrong way" (03/03/2009) | 2 times ØC: sure (3.66) | 3 times ØC: normal (3) |
| "Bank of England ready to pump money into UK economy" (05/03/2009) | 1 time C: sure (4) | |
| "Oil stocks help FTSE 100 regain early losses" (09/03/2009) | 1 time C: normal (3) | |
| "Financial crisis hits world markets" (02/03/2009) | 1 time C: very sure (5) | |
| "FTSE rally fades out as banks head south" (03/03/2009) | | 1 time C: very sure (4) |
| "FTSE edges higher, but Wolseley falls on cash call fears" (03/03/2009) | | 1 time C: normal (3) |
| "International Power leads another FTSE 100 decline" (03/03/2009) | | 1 time C: normal (3) |

## 7. SUMMARY

In this section we sum up the results of the user study.

Participants in Group 1 had no problems to load financial news in the timeline by clicking on a data point in the chart and to understand the connection, although the technique has only been described in the legend. It was deliberately not made aware of the functionality so that users had to discover the legend by their own and try out the new interactive technology. Participants in Group 2 had more difficulties, although the technique was successfully applied in the following tasks. Problems were caused by the allocation of the filter slider to the charts and the resolution of the slider. Because a large period of time is depicted with a relatively small slider, small changes with the slider result in large temporal filtering, and thus provoke a strong period shift in the diagram. The temporal filtering can be controlled with great accuracy only. The problem is systemic in nature, as long periods are filtered with small controllers and cannot be resolved directly. The task was considered with a normal difficulty, with Group 2 required for the task twice as long as Group 1.

Finding appropriate financial news for the price increase was a complex process which resulted in a high average time exposure of 3-6 minutes and a perceived difficulty expressed as difficult. Group 2 took again 1/3 longer to solve the task. Appropriate messages were selected on the first level only by the short title and then had to be cognitively confirmed in the online article. Nevertheless, a majority of users have selected suitable financial news. Table 1 gives an overview of the selected messages. In the independent groups frequently the same news items were chosen as a justification for the price rise and were confirmed with a high confidence level. In many cases the titles already contained an indication for the price increase ("regain early losses", "FTSE rises", "shares bounce back", "lift world markets", "edges higher"), in other cases it must be confirmed by details from the online article. For the second most chosen message "Bank of England ready to pump money into UK economy" it is not intuitively obvious that increasing the money supply leads to an increase of the FTSE index. The online article provides hints like "Bank of England announces that it wants to fight the economic downturn by pumping dog hundreds of billions of pounds into the economy" as a long-term strategy, in the short term the index price decreased: "FTSE dropping 48 points to 3597".

The mapping of the information was perceived as easy to normal difficulty. The users had understood the concept and chose appropriate news items, to connect them with the minimum. The problem was quickly solved by Group 1 (39s), Group 2 in turn took more than three times as long (139s). Table 2 shows that the two most frequently selected messages contain indicators of the index minimum directly in the headline ("worst daily performance", "sends shares wrong way"), but also here the ambiguous news "Bank of England ready to pump money into the UK economy" has been selected.

With the brushing-and-linking technique in Group 1 there were problems due to the small mouse-over zone and optical highlighting, which are based on the large number of data points. In Group 2, this has been improved, but again, the technique was not often recognized as active interaction with the mouse was necessary in order to find the single mapping. The result is that only a part of the users could use the technique and have rated the technique as normal. The technique interaction-icon could be used by almost all participants and was rated as helpful. The specific advantages and disadvantages of both techniques has been worked out independently by the participants: brushing-and-linking is more suitable for many links between the data and shows the correlation or relationship, the interaction-icon acts as a bookmark, and provides a direct link to the associated source and is also suitable for a few connections between the data. Several users suggested applying the techniques in combination, which would combine the advantages. The users felt that the combination of index prices and financial news is useful for the analysis.

## 8. CONCLUSION

The presented search system includes huge amounts of information: price data on the one hand and financial news and online articles on the other. Users can explore this information easily: with one mouse click they can load financial news for a specific day and with another click they can check details in the online article. This way, more information can be scanned, reviewed, compared and judged to find relations between insights from visualizations and context/background information from financial news and online articles. Because a lot of information is accessible on one page, users can compare information more easily and search as well as judge relations. Analog to the approach of dynamic sliders, complex transactions of acquiring and filtering information are hidden behind simple mouse clicks. The user study showed that participants without any domain knowledge could use these interactions techniques and could make decisions with a high confidence which information to relate for patterns in the graph. Comparing the two groups of the user study has shown that researchers could use their world knowledge and experience with information systems to find links between relevant information faster and to confirm with high confidence values. In comparison, the student group took significantly longer and had less confidence, but also similar relations between information items could be confirmed.

**Table 3. Summarized results of the tasks (S=standard deviation, all values are average values)**

| Task | Group 1 | Group 2 |
|---|---|---|
| Self-assessment | good (1, S:0.82) | normal (0.45, S:0.69) |
| 1. Load financial news for index prices | | |
| | 10/11 Time: 64s (S:51s) Difficulty: normal (0.27) | 7/11 Time: 134s (S:106s) Difficulty: normal (0.1) |
| 2. Find reasons for the rise in the market | | |
| Part (a) | 10/11 Confidence: sure (4, S:0.82) | 11/11 Confidence: sure (3.56, S: 0.98) |
| Part (b) | 8/11 Confidence: sure (3.88, S:0.99) | 11/11 Confidence: normal (3.04, S:0.51) |
| | Time: 227s (S:186s) Difficulty: difficult (-0.55) | Time: 360s (S:172s) Difficulty: difficult (-0.82) |
| 3. Link information | | |
| | 11/11 Time: 42s (S:15s) Difficulty: easy (0.64) | 10/11 Time: 139s (S:93s) Difficulty: normal (0.37) |
| 4. Use interaction techniques on the basis of the mapping | | |
| Brushing and Linking | 6/11 Rating: normal (0.33, S:1.58) | 6/11 Rating: normal (-0.09, S: 0.94) |
| Interaction Icon | 10/11 Rating: helpful (0.56, S:1.13) | 9/11 Rating: helpful (0.75, S: 0.65) |
| Overall | Rating: helpful (0.57, S:0.98) | Rating: helpful (0.91, S:0.83) |

Most systems in the area of information visualization are *closed systems*, in the sense that they provide access to information from one or more data sources and show this data in one or more views for the purpose of gaining insight. Information can be explored with several interaction techniques and several views maybe connected by brushing-and-linking. But, these systems have no or limited connections to a network of outer knowledge on the web. However, more and more information and knowledge is available on the web and interlinked like i.e. on web sites, Wikipedia, Web APIs or in the LOD cloud [5]. Interactive visualizations have been brought to the web [10], what makes it possible to create and embed visualizations in web pages, blogs or social networks. But these visualizations are not yet an integrated part of the web and an overall search process, because data and information is hidden behind proprietary software and data holdings and in particular links to related information are missing.

Bates [3] introduces the process of berrypicking that shows that the search process involves browsing over different heterogeneous information to learn from until the satisfactory completion of the task. Marchionini [14] emphasizes in particular the step of learning and investigation in the search process. Card et al. [6] presented the knowledge crystallization loop, in which several steps like information foraging, schema creation, problem solving, learning and decision making are involved. These processes can be more easily coupled if interaction techniques connect resources of the information foraging process, so that the user can easily switch between them.

In this paper we argue for the integration of visualizations in an exploratory search process by the integration of visualizations in the hyperlinking mechanism of the Web. This can be done especially by *integrating links directly in visualizations* like done in this study with the interaction-icon and text links. The user study has shown that *open visualizations* generate an added value. Information in visualizations is not isolated, but uses the link mechanism of the web. The search process will be extended to allow hops between information in visualizations and related resources. Thus, heterogeneous information spaces are connected, which were previously isolated. With the combination the benefits of information visualization and exploratory search can be connected. On the one hand, heterogeneous information is displayed compressed in user-accessible visualizations, on the other hand, they are integrated through links in the search process. The user can browse iteratively over associated resources, can compare, discover connections and gather background knowledge.

In further research we want to apply the presented approach to other domains and information types. For example, statistical information like indicators (GDP, Debts…) can be very well presented in graphs, maps and tables. This information is related to other information items like historical events, news, videos and images that is all available on the web, but not yet integrated accessibly.